\font\mybb=msbm10 at 12pt
\def\bbxx#1{\hbox{\mybb#1}}

\def\C {\bbxx{C}}
\font\mybb=msbm10 at 12pt


\def \aa {\alpha}
\def \bb {\beta}
\def \gg {\gamma}
\def \dd {\delta}
\def \ee {\epsilon}

\def \ll {\lambda}

\def \ss {\sigma}

\def \th {\theta}

\def \ww {\omega}

 \def \ggg {\Gamma}

\def \w {\Omega}

\def \ti {\tilde}
\def \ba {\bar}

\def\bbb{{\bar \beta}}
\def\ggb{{\bar \gamma}}
\def\aab{{\bar \alpha}}

\def\ab{{\bar \alpha}}

\def \2 {{1 \over 2}}
\def \3 {{1 \over 3}}
\def \4 {{1 \over 4}}
\def \5 {{1 \over 5}}
\def \6 {{1 \over 6}}
\def \7 {{1 \over 7}}
\def \8 {{1 \over 8}}
\def \9 {{1 \over 9}}
\def \0 { \infty}

\def\++ {{(+)}}
\def \- {{(-)}}
\def\+-{{(\pm)}}

\def \ek {\eqn\abc$$}

\def \pa {\partial}
\def\na {\nabla}
\def \qq {\qquad}


\def\C{\mkern1mu\raise2.2pt\hbox{$\scriptscriptstyle|$}\mkern-7mu{\rm C}}

\def\i{\iota}
\def\l {\lambda}
\def\a{\alpha}
\def\b{\beta}



\tolerance=10000
\input phyzzx

 \def\unit{\hbox to 3.3pt{\hskip1.3pt \vrule height 7pt width .4pt \hskip.7pt
\vrule height 7.85pt width .4pt \kern-2.4pt
\hrulefill \kern-3pt
\raise 4pt\hbox{\char'40}}}
\def\II{{\unit}}

\def\gij{g_{ij}}

\def\nup#1({Nucl.\ Phys.\  {\bf B#1}\ (}

\def\Dd{{\cal D}}


\REF\coles{ R. Coles \& G. Papadopoulos, {\sl Class. Quantum Grav.}
 {\bf 7} (1990) 427.}
\REF\GPS{G. W.~Gibbons, G.~Papadopoulos and K.S.~Stelle,
{ Nucl. Phys.}\ B {\bf 508}, 623 (1997) hep-th/9706207.}
\REF\gibvan{G.W. Gibbons, R.H. Rietdijk \& J.W. van Holten,
{\sl Nucl. Phys.} {\bf B404} (1993) 42.}
\REF\sfetsos{F. de Jonghe, K. Peeters \& K. Sfetsos,
{\sl Class. Quantum Grav.} {\bf 14} (1997) 35.}
\REF\strom{J. Michelson and A. Strominger, hep-th/9907191.}
\REF\pol{
S. Hellerman and J. Polchinski,
hep-th/9908202.}
\REF\van{C.M. Hull, {\sl Lectures on Nonlinear
Sigma Models and Strings},
 Lectures given in   {\sl Super Field Theories}, eds. H.C.Lee et al, (Plenum,
1987).}
\REF\gates{S.J. Gates, C.M. Hull \& M. Ro\v cek, {\sl Nucl. Phys.}
{\bf B248} (1984) 157.}
\REF\HW{C.M. Hull and E. Witten, Phys. Lett. {\bf 160B} (1985)  398.}
\REF\howepapc{P.S. Howe \& G. Papadopoulos, {\sl Commun. Math. Phys.} {\bf 151}
(1993) 467.}
\REF\howepap{P.S. Howe \& G. Papadopoulos, {\sl Nucl. Phys.} {\bf B289}
(1987) 264; {\sl Class. Quantum Grav.} {\bf 5} (1988) 1647.}
\REF\howepapd{P.S. Howe \& G. Papadopoulos, {\sl Phys. Lett. }
 {\bf B379} (1996) 80.}
\REF\howepapb{P.S. Howe \& G. Papadopoulos, {\sl Nucl .Phys.}
{\bf B381} (1992) 360. }
\REF\DS{
M.\ Dine and N.\ Seiberg,
Phys.\ Lett.\
{\bf B180} (1986) 364.}
\REF\nr{B. deWit and P. van Nieuwenhuizen, Nucl. Phys. {\bf B312} (1989) 58.}
\REF\nra{G.W. Delius, M.Ro{\v c}ek, A. Sevrin and P. van Nieuwenhuizen, Nucl.
Phys. {\bf
B324} (1989) 523.}
\REF\hpb{P.S. Howe and G. Papadopoulos, Commun. Math. Phys. {\bf 151} (1993)
467.}
\REF\alpap{G. Papadopoulos, Nucl.Phys. B448 (1995) 199,
 hep-th/9503063.}
\REF\sigto{ C.\ M.\ Hull, {  Actions for (2,1) Sigma Models and Strings},
Nucl.\ Phys.\ {\bf B509} (1998) 252, [hep-th/9702067].}
\REF\sigtoa{ Mohab Abou Zeid and C.\ M.\ Hull, { The Geometry of sigma models
with twisted supersymmetry},  hep-th/9907046.]}
\REF\lust{E. Kiritsis, C. Counnas \&  D. L\"ust {\sl Int. Journ.
 Mod. Phys.} {\bf A9} (1994) 1361.}
\REF\yano{K. Yano, Differential Geometry on Complex and Almost Complex Spaces,
Pergamon, Oxford,
1965.}
\REF\pop{
C.N.~Pope, M.F.~Sohnius and K.S.~Stelle,
Nucl.\ Phys.\ {\bf B283}, 192 (1987).}
\REF\yanref{K. Yano and M. Ako, Hokkaido Math J. {\bf 1} (1972) 63.}


\Pubnum{ \vbox{ \hbox {QMW-99-16}   \hbox{hep-th/9910028}} }
\pubtype{}
\date{October, 1999}

\titlepage

\title {\bf  The Geometry of Supersymmetric Quantum Mechanics}

\author{C.M. Hull}
\address{Physics Department,
Queen Mary and Westfield College,
\break
Mile End Road, London E1 4NS, U.K.}
\vskip 0.5cm

\abstract {
One-dimensional sigma-models with $N$ supersymmetries are considered.
For conventional supersymmetries there must be $N-1$ complex structures
satisfying a Clifford algebra and the constraints on the target space geometry
can be formulated in terms of these. In the cases in which the complex
structures
are simultaneously integrable, a conventional extended superspace formulation
is given,
with the geometry determined by a 2-form potential for $N=2$, by a 1-form
potential for $N=3$ and a scalar potential for $N=4$; for $N>4$ it is given by
a scalar potential satisfying differential constraints. This gives explicit
constructions of models with $N=3$ but not $N=4$ supersymmetry and of $N=4$
models in which the complex structures do not satisfy a quaternionic algebra.
Generalisations with central terms in the superalgebra are also considered.
 }

\endpage

\chapter{Sigma Models and Supersymmetry}

The conditions for supersymmetry in
one-dimensional sigma-models were given by Coles and Papadopoulos [\coles] and
further studied in [\GPS-\pol].
The analysis is similar to that of
two-dimensional sigma-models,
 for which there is a rich relation between target space
geometry and the amount of supersymmetry. The geometries of the two dimensional
supersymmetric
 models were first classified in
[\van] and have been
   studied extensively  [\van-\lust].
 Remarkably, the conditions in one dimension are considerably weaker, giving a
much wider range of
geometries.
The one-dimensional models have many applications. The moduli spaces for
supersymmetric black holes
are the target spaces of certain $d=1$ supersymmetric sigma-models
and the sigma-model describes the geodesic motion in the moduli space
[\GPS,\strom].
Supersymmetric quantum mechanics  also arises in the  light cone quantization
of  supersymmetric field theories [\pol].
The aim here is to study further  the geometries of one-dimensional
supersymmetric sigma-models
and to give explicit  constructions of certain classes of models.

The standard supersymmetry algebra in one dimension is
$$
\{Q^I, Q^J\}= 2 \delta^{IJ} H
\eqn\qalg
$$
where $\{Q^I; I=1,\dots,N\}$ are the supersymmetry charges and $H$ is the
Hamiltonian.
There are some generalisations, such as the twisted superalgebra
$$
\{Q^I, Q^J\}= 2 \eta^{IJ} H
\eqn\qalgt
$$
for some metric $\eta^{IJ} $ or arbitrary signature [\sigto,\sigtoa], or the
addition of
extra terms
$$
\{Q^I, Q^J\}= 2 \eta^{IJ} H +Z^{IJ}
\eqn\qalgz
$$
The extra generators $Z^{IJ}
$  are central in some cases, and in others their commutators with $Q,H,Z$
lead to further generators and a larger algebra.
The general algebra can be written as
$$
\{Q^I, Q^J\}= X^{IJ}
\eqn\qalgg
$$
and $H$ and $Z^{IJ}$ can then be defined by the trace and trace-free parts of
$X^{IJ}$ with respect to some metric
$\eta^{IJ}$.
In some cases it is more natural to consider \qalgg\ rather than to split $X$
into $H$ and $Z$.
We will consider models in which each of these algebras arise.

The plan of the paper is as follows.
In sections 2,3 and 4, sigma-models with $N=1,N=2$ and $N=4$ supersymmetries
are discussed, reviewing the results of
[\coles,\GPS,\strom] and presenting extended superspace formulations which
give simple derivations of some of the results of [\strom]. For conventional
supersymmetry, there
must be $N-1$ complex structures satisfying certain conditions.
The general $N=2$ geometry is specified by a 2-form potential  (considered
further in section 6).
Generalisations are discussed in which   superalgebras
such as \qalgz\ or \qalgg\ arise, extending the results of [\gibvan,\sfetsos]
to geometries with torsion, allowing
the realisation of $N=2$ supersymmetry on almost complex  manifolds, or on
manifolds with no almost complex structure
(and so of arbitrary dimension and signature) which admit generalised
Yano-Killing tensors.
In section 5, new models with $N=3$ supersymmetry but not $N=4$ supersymmetry
are found, which have three (almost) complex
structures but only two of which lead to extra supersymmetries.
In the case in which the three complex structures are simultaneously
integrable, the geometry is given in terms of  a 1-form
potential, giving the local construction of {\it all} such models.
In section 6, the conditions for     $N$ extended supersymmetry for general $N$
are discussed, requiring the target space to be
a  manifold with a Clifford structures (a set of $N-1$ complex structures
satisfying a Clifford algebra) and to have a geometry
which we refer to as {\it Clifford K\"ahler with Torsion} (CKT), in analogy
with the
nomenclature suggested in [\howepapd]  of
{\it K\"ahler with Torsion} (KT) for the geometry of  (2,0) sigma-models  [\HW]
and
{\it Hyper-K\"ahler with Torsion} (HKT) for the geometry of the (4,0)
sigma-models first
found in [\van].
The case of $N=4$ is considered in section 7. In the special case in which the
three complex structures satisfy
the algebra of unit imaginary quaternions, the  geometry was shown in [\strom]
and in section 4 to be weak HKT.
In the general case (in which the   complex structures do not necessarily
satisfy
the    quaternion algebra), the geometry for the case in which
 the complex structures are simultaneously integrable
 is shown to be
 given locally in terms of   a scalar potential by an expression that reduces
to the expression
of section 4 and [\strom] for the quaternionic case.
For $N>4$, the geometry cannot be weak HKT, and the scalar potential
must satisfy certain differential constraints; non-trivial examples of such
geometries are known for $N=8$ [\GPS].
Under certain conditions
similar to those derived  in [\strom], these models can have $OSp(N/1)$
superconformal symmetry.

\chapter{{\bf  N=1 one-dimensional supersymmetry}}

The simplest form of  N=1 supersymmetric sigma model of [\coles]  is defined on
a   $D$-dimensional manifold ${\cal M}$ with metric $g$ and a 3-form $c$
(which  is not closed in general)
and has an $N=1$ superspace action
$$
I=-{1\over2}\int\, dt d\theta\, \big(ig_{ij}DX^i {d\over dt}X^j+
 {1\over 3!} c_{ijk} DX^i DX^j
DX^k\big)\
\eqn\act
$$
where
$$
D^2=i{d\over dt}\ ,
\eqn\bsix
$$
$t$ is the worldline parameter, $\theta$ is a real fermionic variable and
$X^i(t,\theta)$ is
an unconstrained real superfield, which gives a map from   superspace to
${\cal M}$, with $X^i$
real coordinates on  ${\cal M}$. The generalisation considered in [\coles] in
which fermionic
superfields are added will not be considered here.

Expanding the superfield $X^i$ gives the
 component fields
$$
X^i=X^i|\qquad \lambda^i=DX^i|\ ,
\eqn\bfive
$$
consisting of $D$ scalar fields $X^i$ and $D$ real fermionic fields
$\lambda^i$. The component action is
$$
I={1\over2}\int\, dt\, \big(g_{ij}{d\over dt}X^i\, {d\over dt}X^j+i g_{ij}
\lambda^i\nabla^{(+)}_t\lambda^j
 -{1\over 3!}\partial_{[i}c_{jkl]} \lambda^i\lambda^j\lambda^k\lambda^l\big)
\eqn\btwo
$$
 The
covariant
 derivative
$\nabla^{(+)}_t$ is the pull back of the target space covariant derivative with
torsion $c$
$$
\nabla^{(+)}=\nabla+{1\over2} c
\eqn\bthree
$$
 where
$$
\Gamma^{(+)}{}^i{}_{jk}=\Gamma^i_{jk}+{1\over2} c^i{}_{jk}\ ,
\eqn\bfour
$$
and $\Gamma$ is the Levi-Civita connection of the metric $g$.
  If $c$ is closed then this action   can be
obtained by dimensionally reducing the action  [\HW] of (1,0) supersymmetric
two-dimensional sigma models.

\chapter{{\bf  N=2 one-dimensional supersymmetry}}
\section{Complex Geometry and Supersymmetry}

There are two basic kinds of $N=2$ models in 1 dimension. The $N=2a$ models are
constructed from unconstrained real $N=2$superfields
while the $N=2b$ models use complex chiral superfields. The dimensional
reduction of (1,1) supersymmetric two dimensional sigma-models gives $N=2a$
models while the
reduction of (2,0) supersymmetric two dimensional sigma-models gives $N=2b$
models.
Both have been constructed in [\coles,\GPS]. Here we shall
consider only the N=2b models and the special cases in which they have extra
supersymmetry
(such as the N=4b and N=8b models of [\GPS]) and we shall
 begin by  reviewing the results of   [\coles,\GPS].

To determine
the conditions on the couplings of the action \act\ required by N=2
supersymmetry, we follow [\gates] and express the second
supersymmetry transformation in terms of the N=1 superfield $X$ as
$$
\delta X^i=\eta\, I^i{}_j DX^j\ ,
\eqn\varx
$$
where $\eta$ is the parameter of the transformation.
The $N=2$ superalgebra will be satisfied if
$I$ satisfies
$$
 I^2=-\II
\eqn\alc
$$
and
$$
{\cal N}^k{}_{ij} (I) =0,
\eqn\nij
$$
where ${\cal N}(I)$ is the Nijenhuis tensor of $I$, defined by
$$
{\cal N}^k{}_{ij} (I) \equiv
I^l{}_{i} I^k{}_{[j,l]} -I^l{}_j I^k{}_{[i,l]}
\eqn\comp
$$
The condition \alc\ implies that $I$
 is an almost complex structure, requiring that the target space dimension be
even, and \nij\ implies that the almost complex
structure  is  integrable, and so is a   complex structure.

It was shown  in  [\coles,\GPS]  that the action \act\ is
invariant
 under this
transformation provided   that
$$
\eqalign{
g_{k\ell} I^k{}_i I^\ell{}_j&=g_{ij}
\cr
\nabla^{(+)}_{(i} I^k{}_{j)}&=0
\cr
\partial_{[i}\big(I^m{}_j c_{|m|kl]}\big)-
2 I^m{}_{[i} \partial_{[m} c_{jkl]]}&=0\ .}
\eqn\cons
$$
An alternative derivation  was given in  [\sfetsos].
The first condition is that the
  metric  $g$  is hermitian
with respect to the
complex structure $I$.
The last condition
was written in [\GPS] as
$$
\iota_Idc-{2\over3} d\iota_Ic=0
\eqn\missinga
$$
 where $\iota_I$ is the inner derivation with
respect to the the complex structure $I$.
This acts on an $n$-form $\ww$ as
$$ \eqalign{
\i _I: \ww &= {1\over n!} \ww_{i_1...i_n} dX^{i_1} \wedge dX^{i_2}... \wedge
dX^{i_n}
 \to (\i _I  \ww ),
\cr
(\i _I  \ww ) &= {1\over (n-1)!} \ww_{j[i_2...i_{n-1}}I^j{}_{i_1]} dX^{i_1}
\wedge dX^{i_2}...
\wedge dX^{i_n}
\cr}
\eqn\abc$$

If the metric is hermitian, $c$ is closed and the complex structure is
covariantly  constant
with respect to the $\nabla^{(+)}$ connection,
$$
\nabla^{(+)}_i I^j{}_k=0\ ,
\eqn\actteen
$$
then these conditions
are all satisfied and the model can be obtained by dimensional reduction of a
(2,0) supersymmetric sigma model in two dimensions.
However, these conditions are much stronger than \cons,
so that there are many geometries that allow $d=1, N=2b$ models but not (2,0)
$d=2$ models.

The complex structure enables us to introduce complex coordinates $Z$, so that
$X^i=( Z^\alpha
, \bar Z^{\bar \beta} ) $
($\alpha,\bar \beta=1,...,D/2$) and
the complex structure
is constant
$$I^i{}_j=\pmatrix{
i \delta^\alpha {}_\beta & 0 \cr
0 & -i \delta^{\bar \alpha} {}_{\bar \beta} \cr
}
\eqn\abc$$
  and  the line element for the hermitian metric is
$ds^2=2 g_{\alpha \bar \beta}d Z^\alpha d \bar Z^{\bar \beta}$.

\section{Generalised Symmetries}

For \varx\ to be a symmetry, it is sufficient for \cons\ to be satisfied. If
\cons\ is satisfied but
\alc,\nij\  are not, then the symmetry algebra will not be the usual
supersymmetry algebra \qalg.
For example, if $I^2 =+1$ and \nij\ holds, then $I$ is a real structure and the
algebra is a twisted
superalgebra  \qalgt\ of the type studied in [\sigto,\sigtoa]. If $I$ is an
almost complex structure
satisfying
\alc\ but not
\nij, then the  commutator of two transformations of the form \varx\ gives the
new symmetry
$$\delta X^i= \rho {\cal N}^i_{jk} DX^j DX^k
\eqn\nijvar$$ with bosonic  parameter $\rho$.
 The superalgebra  is then of the form \qalgz\ with $Z^{22} $ the charge
generating this new
symmetry.  This is a central charge commuting with $Q,H$ and so the
 supersymmetry algebra closes without  further generators, by an argument
similar to that given for
two-dimensional sigma models in [\nr - \alpap].

In the general case in which \cons\ is satisfied but
\alc,\nij\  are not, the superalgebra is    of the form \qalgg\ with $X^{11}=H
,X^{12}=0 $, but   $X^{22} $
generates the bosonic symmetry
$$\delta X^i= \rho( 2i R^i{}_j \pa _t X^i+{\cal N}^i_{jk} DX^j DX^k)
\eqn\genvar$$
where
$$
R^i{}_j =( I^2)^i{}_j
\eqn\syan$$
If the trace $( I^2)^i{}_i$ is non-zero, it can be set to one by scaling $I$,
in which case the algebra takes the form
\qalgz\ with $Z^{11}=H ,Z^{12}=0 $, but   $Z^{22} $
generating the symmetry  \genvar\ with $R^i{}_j =( I^2-\II)^i{}_j
$.
Such symmetries have been considered in [\gibvan,\sfetsos].

\section{The Torsion-Free Case}

 If
the torsion $c$ vanishes, then the second condition in \cons\ becomes  the
condition
$$
\nabla _{(i} I^k{}_{j)} =0
\eqn\tach$$
implying
that $I_{ij}$ is a
  Yano Killing-tensor, as was   pointed out in the context of one-dimensional
supersymmetric sigma models in [\gibvan].
For an almost complex structure satisfying \tach,   the hermiticity condition
in \cons\ together with \alc\
implies that $( {\cal M},g,I)$ is an almost Tachibana space [\yano].
The condition \tach\
implies that the Nijenhuis tensor can be written as
$${\cal N}_{ij}^k=-I^k{}_l \nabla _i I^l{}_j
\eqn\ffsfdf$$
Then if in addition $I$ is a complex structure with ${\cal N}(I)=0$,  \ffsfdf\
implies that the complex
structure must be covariantly constant, $\nabla I=0$, and the space must be
K\"ahler.  (An almost
Tachibana space with vanishing  Nijenhuis tensor is also called a  Tachibana
space, but  this is the
equivalent to the K\"ahler condition.)

If $I $ is not an almost complex structure, but is a general  Yano
Killing-tensor, then the
tensor $R^i{}_j =( I^2)^i{}_j$
is a Stackel-Killing tensor and the symmetry algebra is of the form \qalgg\
with  $X^{22} $
generating the symmetry \genvar. This is the case analysed  in [\gibvan], where
a number of examples were considered.
  Models with $N=2$ supersymmetry of this type
can arise for odd dimensional target spaces as well as for even dimensional
ones, and for
Lorentzian signature target spaces,
such as the Kerr-Newman black hole [\gibvan]. It would be interesting to
investigate  whether there are BPS states
associated with the central charge   $Z^{22} $.

\section{N=2 Superspace Action}

The $N=2b$ one-dimensional supersymmetric model can be written in $N=2 $
superspace with coordinates $t,\th ^0, \th ^1$
and supercovariant derivatives
$D_0, D_1$ satisfying
$$
D_0^2=i{d\over dt}\ ,
\qquad
D_1^2=i{d\over dt}\ ,
\qquad
D_0D_1+D_1D_1=0\ .
\eqn\beight
$$
It is useful to define
$\Dd=D_0+iD_1$, $\th = \th ^0 +i \th ^1$
so that
$$\Dd ^2=0, \qq \{ \Dd, \ba \Dd \}= 2i{d\over dt}\ ,
\eqn\salg$$
and introduce
  chiral  superfields $Z^\aa$ and their complex conjugates $\bar Z^{\bar \aa}$
satisfying
$$
\bar \Dd Z^\a=0
\eqn\chir$$
The most general superspace action
is given by
$$
I={1\over 4}\int dt\, d^2\theta\, \big(iG_{\a \bbb} \Dd Z^\a
\ba \Dd \ba Z^{\bbb}
+{1\over 2} B_{\a \b}\Dd Z^\a \Dd Z^\b
+{1\over 2}\ba B_{\aab \bbb}\ba \Dd \ba Z^\aab \ba \Dd \ba Z^\bbb
\big)\ ,
\eqn\actt
$$
for some  $G_{\a \bbb},B_{\a \b}$, with $\ba B_{\aab \bbb}=(B_{\a \b})^*$,
plus a chiral superpotential term
$$
S= \int dt\, d \theta\,W(Z) + \int dt\, d \ba \theta\,\ba W(\ba Z)
\eqn\pot
$$
for some holomorphic function $W$.
The action is invariant under
$$
B_{\a \b} \to B_{\a \b} + \pa _{[\a } \ll _{\bb]}
\eqn\kagag$$
and so only depends on the holomorphic field strength
$$e_{\a\b \gg}= 3\pa _{[\a } B_{\b\gg]}
\eqn\abc$$

Dimensional reduction of the superspace action [\DS] for the (2,0) model in two
dimensions [\HW] gives an action of the form
$$
I={i\over 2}\int dt\, d^2\theta\, \big(k_\a \pa _t Z^\a - \ba k_\aab \pa _t \ba
 Z^\aab
\big)\ ,
\eqn\twofdfd
$$
where $k_\a $ is the potential introduced in [\HW],
but this can be rewritten using \salg,\chir\ as
$$
I={1\over 4}\int dt\, d^2\theta\, \big( k_{\alpha, \bar \beta}+ k_{\bar \beta ,
\alpha}
\big)\Dd Z^\a
\ba \Dd \ba Z^{\bbb}
\ ,
\eqn\twofdd
$$
which is of the same form as the $G_{\a \bbb}$ term in \actt,
and so   terms of the form \twofdfd\
are already included, and \actt\ is indeed the most general action.

The $N=2$  superfields $Z^\a$  give the $N=1$ superfields  ($D \equiv D_0$)
$$
Z^\a=Z^\a|_{\th^1=0}
\qquad D Z^\a=D_0 Z^\a={1\over 2} \Dd Z^\a|_{\th^1=0}
\
\eqn\dfgfjk
$$
Then the $\th^1 $ integral in   \actt\ can be performed (see appendix for
details) to give the $N=1 $
superspace action \act, with
$$g_{\a\b}=0, \qq g_{\a\bbb}=G _{\a\bbb}
\eqn\abc$$
and
$$
c_{\a\b
\ggb}=  2(g_{\a\ggb ,\b}-g_{\b\ggb,\a}), \qq c_{\a\b\gg}=12i \pa _{[\a }
B_{\b\gg]}
\eqn\abc$$
together with the complex conjugate equations.
Thus the geometry is completely specified in terms of
(i) a hermitian metric $g_{\a\bbb}$ and (ii) a   (3,0) form
$e$  satisfying $  \pa e=0$, so that it can be expressed locally as
$e=\pa B$ for some (2,0) form potential $B$.
Then, in terms of the holomorphic exterior derivative  $\pa$ with $d= \pa + \ba
\pa $,
$$c= {i }(\pa - \ba \pa ) \ww +4i(e -\ba e )
\eqn\abc$$
where
$\ww(I)$ is the fundamental form constructed from $I$:
$$\ww(I)= {1\over 2} I_{ij} dX^i \wedge dX^j = i g_{\a \bbb} dz^\a \wedge d\ba
z ^\bbb
\eqn\abc$$
If $e=0$, then for a given hermitian metric, $c= {i }(\pa - \ba \pa ) \ww$ is
the unique torsion
3-form such that the complex structure is covariantly constant,
$$
\na _i ^{(+)}I ^j{}_k=0
\eqn\Iconi
$$
In the case in which $I$ is an almost complex structure,
the unique torsion three-form
for which the complex structure is covariantly constant,
 \Iconi, is
$$
c_{ijk}=4N_{ijk}+ I_{[mn,p]}I^m{}_iI^n{}_jI^k{}_p
\eqn\fghkjh$$
Examples of $d=1,N=2$ supersymmetric models on almost complex manifolds
with $c$ given by \fghkjh\
arise from the dimensional reduction of the models of [\nr -
\alpap].

\chapter{{\bf  N=4 one-dimensional supersymmetry}}
\section{Conditions for $N=4$ Supersymmetry}

One-dimensional N=4b
supersymmetric sigma models  arise from the dimensional
reduction of the two-dimensional (4,0)
supersymmetric sigma-model. The geometry of these $D=2$ models was first found
in [\van],
and the name hyper-K\"ahler with torsion (HKT) has been proposed for this
geometry  [\howepapd].
  The geometry associated with the N=4b
model in 1 dimension
 is not necessarily HKT, but satisfies weaker conditions [\coles,\GPS].
The extended supersymmetry
transformations can be written in terms of  N=1 superfields as [\gates]
 $$
\delta X^i=\eta^r I_r{}^i{}_j DX^j
\eqn\abone
$$
where $\{\eta^r; r=1,2,3\}$ are the supersymmetry parameters and
$\{I_r; r=1,2,3\}$ are
tensors on {\cal M}.
The conditions from the closure of the N=4 supersymmetry algebra are [\coles]
$$
\eqalign{
I_r I_s+I_s I_r&=-2 \delta_{rs} \II
\cr
{\cal N}(I_r, I_s)&=0}
\eqn\abtwo
$$
where ${\cal N}(I_r, I_s)$ is the Nijenhuis tensor for the pair  $(I_r, I_s)$,
so that the $I_r$ are three
complex structures that anti-commute with one another. The conditions for the
invariance of the
action are [\coles,\GPS]
$$
\eqalign{
g_{k\ell} I_r{}^k{}_i I_r{}^\ell{}_j&=g_{ij}
\cr
\nabla^{(+)}_{(i} I_r{}^k{}_{j)}&=0
\cr
\iota_rdc-{2\over3}d\iota_rc&=0\ ,}
\eqn\conf
$$
where $\iota_r$
denotes inner derivation with respect to the complex structure $I_r$ [\GPS].
The metric is hermitian with
respect to all complex structures.

 A {\it
weak} HKT manifold is a Riemannian manifold
$\{{\cal M}, g , c\}$ equipped with a metric $g$, a three-form $c$ and three
complex structures
$\{I_r; r=1,2,3\}$ satisfying the algebra of imaginary unit quaternions
$$
I_r I_s=- \delta_{rs}+\epsilon_{rst} I_t\ ,
\eqn\quat
$$
such that the metric is hermitian with respect to all complex structures
and the complex structures are each covariantly constant with
	respect to the
$\nabla^{(+)}$ covariant derivative
$$
\nabla^{(+)}_k I_r{}^i{}_j=0\ .
\eqn\abfive
$$
\noindent If in addition the three-form $c$ is closed, then ${\cal M}$ has a
{\it strong} HKT structure.
The target space of
two-dimensional (4,0)-supersymmetric sigma models has a strong HKT
structure
[\van] while
  any weak  HKT manifold   solves all the
conditions required by N=4b one-dimensional supersymmetry [\GPS].
It was argued in [\strom] that the conditions for N=4b one-dimensional
supersymmetry
with \quat\
are equivalent to the ones for weak HKT geometry, but  in [\GPS]  examples are
given of   models
admitting N=4b   supersymmetry (and in fact N=8b supersymmetry) but which are
not weak HKT.
 We will
now investigate further  the general solution of the N=4b   supersymmetry
conditions, using
superspace constructions, and aim to clarify the relation between weak HKT
geometries and the $N=4$
supersymmetry conditions.

\section{N=2 Superspace Formulation}

Suppose the complex structures satisfy the quaternion algebra \quat.
One of the extra supersymmetries, that parameterised by $\eta_3$, say, can be
made manifest by using an $N=2$ superspace
formulation with chiral superfields, so that the action  is of the form  \actt.
The remaining two  supersymmetry transformations can be written as
$$ \delta Z^\alpha =  \eta J^ \alpha {}_ \bbb \bar \Dd \bar Z^ \bbb
\eqn\varz$$
together with the complex conjugate relation,
where $J= {1 \over 2} (I_1-iI_2)$ and $   \eta= \eta_1 +i \eta_2$.
This is consistent with the chirality constraint
$\Dd Z=0$ provided $I_r$ are complex structures and
$$ J^ \alpha {}_ {[\bbb, \bar \gamma]}=0
\eqn\jcon$$
which is the condition that the Nijenhuis tensor
${\cal N} (I_r,I_3)=0$ for $r=1,2$.
This then implies  that the remaining
Nijenhuis tensors also vanish [\strom,\yanref].
The condition \jcon\ implies that locally there is some $f^\alpha$ such that
\varz\ can be written as
$$ \delta Z^\alpha =  \eta  \bar \Dd f^\alpha
\eqn\abc$$
with
$$J^ \alpha {}_ {\bbb}=\pa _\bbb f^\alpha
 \eqn\abc$$

The terms in the variation of the action \actt\ involving $\eta$ are
$$\eqalign{
\dd I=&
{1\over 4}\int dt\, d^2\theta\, \eta \Big(
-{i \over 2}
\hat \na _\a J_{\bbb \ggb}
 \Dd Z^\a
\ba \Dd \ba Z^{\bbb}\ba \Dd \ba Z^{\ggb}
\cr &
+2 J_{(\ab \bbb)} \pa _t \ba Z^\ab \ba \Dd \ba Z ^\bbb
\cr &
+{1\over 2} e_{\a \b \gg}J^\gg{}_\ggb \Dd Z^\a \Dd Z^\b \Dd \ba Z^{\ggb}
-{1\over 6}\ba e_{\aab \bbb \ggb ,\a}f^\a
\ba \Dd \ba Z^\aab \ba \Dd \ba Z^\bbb \ba \Dd \ba Z^{\ggb}
\Big)\ ,\cr}
\eqn\transi
$$
where $\hat \nabla$ is the covariant derivative
preserving $I_3$, $\hat \na I_3=0$, so that the    torsion  3-form for the
connection $\hat \na $
is given by
$ i(\pa - \ba \pa)\ww_3$, where $\ww_3$ is the   two-form constructed from
 $I_3$.
The terms involving $\ba \eta$ are obtained by complex conjugation.
The terms involving $\eta$ must vanish separately from those
involving
$\ba \eta$, which requires
$$\eqalign{
&J_{(\a  \b )}=0
\cr &e=0
\cr & \hat \nabla _iI_r{}^j{}_k =0
\cr}
 \eqn\conc$$
so that $\na ^{(+)}=\hat \na $ and the space is weak HKT. The same result was
obtained studying the conditions \conf\
 in complex
coordinates in [\strom]. Thus the only $N=4b$ models for which the complex
structures satisfy the
quaternion algebra are those with weak HKT target spaces.

\section{N=4 Superspace}

The   one-dimensional N=4b supersymmetry multiplet can be written in N=4
superspace   with
coordinates
 $\{t, \theta^0, \theta^r;
r=1,2,3\}$ and the constraints
$$
D_rX^i=I_r{}^i{}_j D_0X^j
\eqn\absix
$$
The supersymmetry derivatives satisfy the algebra
$$
\eqalign{
D_0^2&=i{d\over dt}
\cr
D_0D_r+D_r D_0&=0
\cr
D_sD_r+D_r D_s&=2i\delta_{rs}{d\over dt}\ .}
\eqn\abseven
$$
The action
$$
I=-{1\over2}\int\, dt d\theta^0\, \big(i g_{ij}D_0X^i {d\over dt}X^j+
 {1\over 3!} c_{ijk} D_0X^i
D_0X^j D_0X^k\big)\
\eqn\abeight
$$
was given in [\GPS].

A more useful action can be
given for the case in which  the three complex
structures are simultaneously integrable, that is, there is a local coordinate
choice
for which all three complex structures have constant components.
The construction is very similar to that of [\gates].
Here we will discuss the special case in which the complex structures satisfy
the quaternion
algebra \quat, and will defer the general case until section 7.3. It is
convenient to use two complex fermionic
superspace coordinates
$\th ^a$,
$a=1,2$, instead of four real ones, with the supercovariant derivatives $\Dd_a
$ and their complex conjugates
$\ba
\Dd^a$ satisfying
$$\{ \Dd_a , \Dd _b \}=0, \qq \{ \Dd_a , \ba \Dd ^b \}=2i \pa _t \dd_a {}^b
 \eqn\abc$$
Choosing a coordinate system in which the complex structures satisfying \quat\
take the convenient
form
$$ I
_1=\pmatrix{
i & 0 \cr
0 &-i \cr
} \otimes \II,
\qquad
I_2=\pmatrix{
0 & \sigma_2  \cr
-\sigma_2 & 0 \cr
}\otimes \II
,
\qquad
I_3=\pmatrix{
0 & i\sigma_2  \cr
i\sigma_2 & 0 \cr
}\otimes \II
 \eqn\abc$$
where $\II$ is the $n\times n$ identity matrix,
 the constraints can be written as
follows. The complex dimension must be even, and the fields $Z^\a$
($\a=1,...,2n$) split into two sets $z^A, w^A$
where  $A=1,...,n$ and satisfy the constraints
$$\eqalign{
&\ba \Dd^a z^A=0,
 \qq
\ba \Dd^a w^A=0
\cr &
\Dd _2 w^A =-i \ba \Dd^1 \ba z ^A, \qq
\Dd _2 z^A =i \ba \Dd^1 \ba w ^A
\cr}
 \eqn\chirf$$
These are a truncation of the $N=4$ twisted chiral constraints of [\gates]
(given by restricting to negative chirality).

The general N=4 superspace
action for this twisted chiral N=4b multiplet is
then
$$
I={1\over 4}\int dt\, d^4\theta\, L(z,\ba z, w, \ba w)
 \eqn\actf$$
for an arbitrary function $L$.
Integrating over $\th^2, \ba \th _2$ gives an N=2 superspace action $\Dd _2 \ba
\Dd_2 L$ which can be rewritten using
\chirf\ to be of the
form \actt\ with
$$ e_{ijk}=0
 \eqn\abc$$
and metric $g_{\a \bbb}=G_{\a \bbb}$ given by
$$
\eqalign {
g_{z^A \ba z^B}&= \pa_{w^B}\pa _{\ba w^A} L + \pa_{z^A}\pa _{\ba z^B} L
\cr
g_{w^A \ba w^B}&= \pa_{z^B}\pa _{\ba z^A} L + \pa_{w^A}\pa _{\ba w^B} L
\cr
g_{z^A \ba w^B}&=
 -\pa_{z^B}\pa _{\ba w^A} L + \pa_{z^A}\pa _{\ba w^B} L
\cr
g_{w^A \ba z^B}&= \pa_{w^A}\pa _{\ba z^B} L - \pa_{z^A}\pa _{  w^B} L
\cr}
 \eqn\metf$$
giving a simple superspace derivation of the result  of [\strom].
We have recovered the result of [\strom] that the geometry of any weak HKT
space with simultaneously integrable complex
structures can be given locally in terms of a potential $L$ by \metf.
However, many HKT spaces do not  have simultaneously integrable complex
structures (e.g. any non-trivial hyper-K\"ahler space)
and in those cases the geometry  is not
given by an unconstrained potential in general (e.g. for
hyper-K\"ahler spaces, the geometry is given by a K\"ahler potential satisfying
highly non-trivial constraints).

It is perhaps worth noting that the conditions (given by equation (38) of
[\gates])  for a function $ L(z,\ba z, w, \ba w)$ to
determine a two-dimensional  off-shell (4,4) supersymmetric  sigma-model
constructed from twisted chiral multiplets are
precisely the conditions  that the
metric  $g_{\a \bbb}$ defined by \metf\ vanish.

Introducing the notation $Z^{Au}=\{ z^A, w^A \}$ where $u=1,2$ so that $Z^{A1}=
 z^A $
and $Z^{A2}=  w^A $,
and the  complex conjugate $\ba Z^A{}_u$,\foot{Complex conjugation raises or
lowers the indices $a,b$ and $u,v$ but not the real indices $A,B$.}  the
constraints \chirf\
can be written as
$$
\ba \Dd ^a Z^{Au}=0, \qq
\Dd _a Z^{Au}=-i \ee ^{ab} \ee ^{uv} \ba \Dd ^b \ba Z^A{}_v
  \eqn\abc$$
This is the form of the constraint used in [\pol], where the   action \actf\
was also considered.
The metric \metf\ can be rewritten as
$$
g_{Au \, B }{}^v =
{\pa ^2 L\over \pa Z^{Cw} \pa \ba Z^C{}_x}
\left(
\dd^C{}_A \dd^D{}_B\dd^w{}_u \dd^v{}_x
-\dd^D{}_A \dd^C{}_B(
\dd^v{}_u \dd^w{}_x
-
\dd^w{}_u \dd^v{}_x )
\right)
 \eqn\abc$$

\chapter{{\bf  N=3 one-dimensional supersymmetry}}
\section{Complex Structures and Supersymmetry}

We now return to the   relation between the argument of [\strom] that N=4
supersymmetry (with
\quat) implies weak HKT geometry and the explicit examples of models in [\GPS]
with $N \ge 4$
supersymmetry whose geometry is {\it not} weak HKT, but is in fact  a geometry
with 7 complex structures that was
termed
OKT in [\GPS] (see section 7). The resolution lies in an important  difference
between the supersymmetric sigma model geometries in 1 dimension  [\coles].
In both cases, there are $m$ complex structures satisfying a Clifford algebra
$$I_r I_s+I_s I_r=-2 \delta_{rs}
 \eqn\abc$$
where $r=1,...,m$, giving the possibility of $N=m+1$ supersymmetries in $D=1$
and $(m+1,0)$
supersymmetry in $D=2$.
If the complex structures are each covariantly constant with respect to some
connection $
\hat \nabla$ (possibly with torsion)
$$
\hat \nabla I_r=0
 \eqn\abc$$
then the complex structures must commute with the holonomy group of this
connection, and if the holonomy is irreducible this
implies that the complex structures must form a division algebra, so that the
only possibilities
are $m=0,m=1$ and $m=3$ (the octonion algebra  cannot be represented by a set
of matrices).
Thus in $D=2$, the only $(N,0)$ supersymmetries of standard type that can arise
 for rigid supersymmetric sigma-models are
those for
$N=1,2,4$ [\van]. (Other possibilities such as (3,0) can arise for the more
general sigma-models of the type considered in [\van].)
In particular, given two covariantly constant complex structures $I_1,I_2$,
their product $I_3=I_1I_2$ must also be
a  covariantly constant complex structure and (3,0) supersymmetry implies
(4,0).
However, in $D=1$ the complex structures are not necessarily covariantly
constant, but satisfy the weaker condition
$$\nabla^{(+)}_{(i} I^k{}_{j)} =0
\ek
Consider the case of
 $N=3$ models with two    complex structures $I_1,I_2$ satisfying \abtwo,\conf.
The product
$ I_3= I_1I_2$ is an almost complex structure
but in general it will not
 satisfy the supersymmetry conditions \conf, so that
$N=3$ does not necessarily imply $N=4$  supersymmetry [\coles].
However, if $I_3= I_1I_2$ is a complex structure satisfying these constraints,
then the  $I_r$ satisfy the quaternion algebra
\quat\ and
the geometry must be weak HKT, as we saw in the last section.
In particular, for the N=8 supersymmetric models of [\GPS], the 7 complex
structures
satisfy a Clifford algebra but not a division algebra, so that  the product  of
any  two complex structures is not a complex
structure corresponding to   a supersymmetry, and so the OKT geometries need
not be weak HKT,
 and none of the OKT examples in
[\GPS] are.
More generally, a target space which is not
weak HKT
can have $N \ge 3 $ supersymmetry provided that for any two complex structures
$I_1,I_2$ that
correspond  to supersymmetries, the product  $I_1I_2$ does not
lead to a supersymmetry.

In this section we will investigate $N=3b$ models in $D=1$ further. These are
of a different type to   the
$N=3$ models constructed in [\coles], which are based on a real $N=2$
supermultiplet, and require the target space structure
group to be reducible. We require two complex structures $I_1,I_2$ satisfying
\abtwo,\conf.
The supersymmetry corresponding to $I_1$ can be made manifest by using $N=2$
superspace with chiral superfields and
action \actt.
 The extra supersymmetry transformation corresponding to $J=I_2$ can be written
as
$$ \delta Z^\alpha =  \eta J^ \alpha {}_ \bbb \bar \Dd \bar Z^ \bbb
\eqn\vart$$
which is of the same form as
\varz, but now with the important difference that $
\eta$ is the {\it real } parameter
corresponding to the third supersymmetry, whereas $\eta$ was the complex
parameter
$   \eta= \eta_1 +i \eta_2$ in \varz. Moreover, $J=I_2$ here, whereas in \varz\
we had
$J= {1 \over 2} (I_1-iI_2)$.
Again
$$ J^ \alpha {}_ {[\bbb, \bar \gamma]}=0
\eqn\abc$$
so that locally there is some $f^\alpha$ such that
\vart\ can be written as
$$ \delta Z^\alpha =  \eta  \bar \Dd f^\alpha
\eqn\abc$$
with $$J^ \alpha {}_ {\bbb}=\pa _\bbb f^\alpha
 \eqn\abc$$
The conditions for invariance  of the action \actt\ are
$$\eqalign{&
J_{(\a \b )}=0
\cr &
\hat \na _\ab J_{\a \b    }
=
 e_{\a \b \gg}J^\gg{}_\ab
\cr &\ba e_{\aab \bbb \ggb ,\a}=0
\cr}
\eqn\conth$$
These are found from requiring that the sum of \transi\ and its complex
conjugate vanish for real $\eta$.
Thus there will be $N=3$ supersymmetry for any hermitian manifold with (3,0)
form $e=\pa B$ and an extra complex structure $J$
provided the conditions \conth\ are satisfied.
We will now construct a large class of $N=3$ models satisfying these
constraints
and which are not $N=4$ supersymmetric.

\section {N=3 Superspace Construction}

In the special case in which the two complex structures $I_1,I_2$
are simultaneously integrable, we can use a superspace formulation similar to
that used in section 4.3.
The supercovariant derivatives can be taken to be
$\Dd, \bar \Dd, \ti D$ with $\ti D$ real, satisfying
$$
\eqalign{&
\{ \Dd  , \Dd   \}=0, \qq \{ \Dd  , \ba \Dd  \}=2i \pa _t
,
\cr &
\{ \ti D , \Dd   \}=0, \qq \ti D^2=i \pa _t
\cr}
 \eqn\abc$$
As in section 4.3, we take the complex dimension of the target space to be be
even, and the fields
$Z^\a$ ($\a=1,...,2n$) split into two sets
$z^A, w^A$ where  $A=1,...,n$. If the complex structures are taken to be of the
form
$$ I
_1=\pmatrix{
i & 0 \cr
0 &-i \cr
} \otimes \II,
 \qquad
I_2=\pmatrix{
0 & i\sigma_2  \cr
i\sigma_2 & 0 \cr
} \otimes \II
 \eqn\abc$$
 where $\II$ is the $n\times n$ identity matrix, then the constraints are
$$\eqalign{
&\ba \Dd  z^A=0,
 \qq
\ba \Dd  w^A=0
\cr &
\ti D w^A =-{1\over \sqrt 2} \ba \Dd  \ba z ^A, \qq
\ti D z^A ={1\over \sqrt 2} \ba \Dd \ba w ^A
\cr}
 \eqn\abc$$
which can be rewritten in terms of $Z^{Au}=\{ z^A, w^A \}$ where $u=1,2$,
$Z^{A1}=  z^A $
and $Z^{A2}=  w^A $,
 as
$$
\ba \Dd ^a Z^{Au}=0, \qq
\ti D Z^{Au}= {1\over \sqrt 2}   \ee ^{uv} \ba \Dd   \ba Z^A{}_v
  \eqn\chiru$$
The general action can be written in terms of an unconstrained 1-form potential
 $k_i(X)=(k_{Au} , \ba k_A{} ^u )
$ as
$$
I=-{1\over 2}\int dt\, d^3\theta\, \left( k_{Au} \Dd Z^{Au} + \ba k_A{} ^u \ba
\Dd \ba  Z^A_u
\right)
 \eqn\acth$$
Note that a term of the form $ h_{Au} \ti DZ^{Au} $
could be rewritten using the constraint \chiru\ to be proportional to $\ba \Dd
\ba Z$ instead of
$\ti DZ$, and so can be absorbed into the
$\ba k$ term in \acth.

Integrating over the third $\th $
gives an $N=2$ superspace action of the form \actt\ with
$$
g_{Au \, B }{}^v = {1\over \sqrt 2}
\left(
\ee^{wv}
\left[
{\pa k_{Bw}
\over \pa Z^{Au}}
-
{\pa k_{Au}
\over \pa Z^{Bw}}
\right]
+
\ee_{wu}
\left[
{\pa \ba  k_{B }{}^v
\over \pa \ba  Z^{A }_w}
-
{\pa \ba  k_{A }{}^w
\over \pa \ba  Z^{B }_v}
\right]
\right)
 \eqn\abc$$
and
$$
B_{Au\, Bv}= {1\over \sqrt 2}
\left[\ee_{wu}
{\pa    k_{Bv }
\over \pa \ba  Z^{A }_w}
-
\ee_{wv}
{\pa    k_{Au }
\over \pa \ba  Z^{B }_w}
\right]
 \eqn\abc$$
This gives the general construction of $N=3$ models with two simultaneously
integrable complex structures
in terms of a single potential $k$, and in the general case with $B\ne 0$
the complex structures will not be covariantly constant but the weaker
conditions \conth\ will be
satisfied. If $B= 0$, then the complex structures are   covariantly constant
and their product
will   be  a third covariantly
 constant complex structure and
the space will be weak HKT, with N=4 supersymmetry.

\chapter{{\bf   One-dimensional supersymmetry for general N }}
\section{Clifford Structures and Supersymmetry}

In this section we will consider sigma models with $N$ supersymmetries for any
$N$, so that
the $N=1$ model \act\ is   invariant under an additional $N-1$ supersymmetry
 transformations
$$
\delta X^i =\eta^a I_a{}^i{}_j DX^j
\eqn\varn
$$
where $\{ \eta^a; a=1,\dots, N-1\}$ are the supersymmetry parameters.
The conditions required by
the closure of the supersymmetry algebra are
$$
I_a I_b+I_b I_a=-2 \delta_{ab} \II
\eqn\clif
$$
and
$$
N(I_a, I_b)=0
\eqn\nab
$$
and the conditions required by the invariance of the action are
$$
\eqalign{
g_{k\ell} I_a{}^k{}_i I_a{}^\ell{}_j&=g_{ij}
\cr
\nabla^{(+)}_{(i} I_a{}^k{}_{j)}&=0
\cr
\iota_adc-{2\over3}d\iota_ac&=0\ ,}
\eqn\conn
$$
where $\iota_a$ denotes inner derivation with respect to  $I_a$.
We shall call a set of $m$ complex structures satisfying
 \nab\ a {\it Clifford Structure},
and call a Riemannian manifold $\{{\cal M}, g, c\}$
equipped with metric $g$, antisymmetric tensor $c$, and complex
structures $\{I_a\}$ that obey the compatibility conditions \nab\
and \conn\ a  {\it Clifford K\"ahler with Torsion} manifold, or CKT
for short.
The name
 {\it Octonionic K\"ahler with Torsion}   (OKT) was suggested in [\GPS] for
the special case in which $m=7$.

The more general case in which the first condition in \clif\ is satisfied but
\nab\
is not, so that the $I_r$ are almost complex structures, will lead to an
enlarged supersymmetry algebra of the form
\qalgz. We will refer to geometries satisfying \clif,\conn\ but not \nab\ as
{\it Almost
Clifford K\"ahler with Torsion} manifold,
or ACKT for short.
This constitutes a generalisation of the almost Tachibana spaces that arose in
section 3.2.

More generally, if neither  \clif\ nor \nab\ is satisfied but \conn\ holds,
then
the
  superalgebra is    of the form \qalgg\ with $X^{00}=H ,X^{0r}=0 $, but
$X^{rs} $
generates the bosonic symmetry
$$\delta X^i= \rho \left( 2i (R_{rs})^i{}_j \pa _t X^i+{\cal N}(I_r,I_s)^i_{jk}
DX^j DX^k \right)
\eqn\genvara$$
where
$$
 R_{rs}  = {1 \over 2}\{ I_r, I_s\}  \eqn\syans$$
We will restrict ourselves to the CKT case in what follows.

Which values of $N$ can arise?
In [\GPS], models with $N=0,1,2,4,8$ were constructed, and in section 5 models
with $N=3$ were found.
Supersymmetry transformations of the form \varn\ satisfying \clif\ can be found
for any $N$. For example,
consider the case in which all complex structures are constant matrices in some
coordinate system.
Then the complex structures satisfy a Clifford algebra and can be realised as
gamma matrices, which must be real.
If the target space dimension is $D=2^{d/2}$, then the condition \nab\ is
satisfied by $d+1$ complex gamma matrices
$ (\gg_a )^i{}_j$ satisfying the Clifford algebra  corresponding to $O(d+1)$.
If $m$ is the number of these that can be chosen to be simultaneously real,
then these can be used to construct a realisation
of $N=m+1$ extended supersymmetry. For example, for $D=2$ there are 3 complex
gamma matrices
satisfying
\clif, which can be taken to be
$\gg _a = i \ss _a$ for $a=1,2,3$, but of these only $i \ss_2$ is real, so that
$m=1$ and only $N=2$
supersymmetry is possible.
 For $D=2$, $m=1$, for $D=4$, $m=3$,  for  $D=8$, $m=7$ and it is clear that
$m$ can be made arbitrarily large by taking  $D$   large enough.
There are   manifolds admitting Clifford structures for arbitrarily large
values of $m$.

Supersymmetry with any $N$ can be realised on flat space with $c=0$
and $g_{ij}=\dd_{ij}$. For given $D$,
there are  $m$ real gamma matrices, and the transformation corresponding to a
given one
will preserve the free action provided \conn\ are satisfied with
$c=0$ and $g_{ij}=\dd_{ij}$, which will be the case if
the corresponding  gamma matrix is   anti-symmetric. For general geometries and
general Clifford structure, the conditions become more and more restrictive the
higher the value of
$N$. There are non-trivial examples for $N=8$ [\GPS], but it seems likely that
for high enough $N$ the
geometry will be required to be trivial. The $N=16$ models considered in [\GPS]
that arise for black hole moduli spaces     have flat target spaces.

Given $m=N-1$ almost complex structures satisfying \clif,
the products
$$I_{rs} \equiv I_r I_s = {1 \over 2} [I_r,I_s] = -I_{sr}
\ek
are also almost complex structures, and
further tensors
can be formed by taking anti-symmetrised products:
  $$I_{rs...t} \equiv I_r I_s...I_t = I_{[rs...t]}
\ek
The tensors $I_{r_1...r_n}$  (with $n \le m$) are   almost complex structures
($I^2=-\II$) for $n=4k+1,4k+2$
and are
almost real structures or almost product structures
($I^2=\II$) for $n=4k ,4k+3$, where $k=1,2,3,...$. These generate the
enveloping algebra of the Clifford algebra.
Note that the set of all almost complex
complex structures constructed in this way do not satisfy a Clifford algebra in
general;
for example,
$$[I_r, I_{st}]=2 I_s \delta _{rt} -2 I_t \delta_{rs}
\ek

Consider the case of $m=3$.
There are three almost complex structures $I,J,K$ say, which anti-commute with
each other and which each squares  to $-\II$.
If $IJ=K$, then they satisfy the quaternion algebra \quat. If not, then we can
define the products
$$\tilde K=IJ, \qquad \tilde I= JK , \qquad \tilde J= KI
\ek
each of which is an almost complex structure.
Moreover,
$$ [I, \tilde I]=0, \qquad  [J, \tilde J]=0, \qquad  [K, \tilde K]=0
\ek
and there are several subalgebras which are isomorphic to the quaternion
algebra \quat, such as those generated by
$(I,J, \tilde K)$ or   $(\tilde I,\tilde J, \tilde K)$. The $\ti I$ will play a
role in the following section.

In the case in which there are $n$ complex structures which are simultaneously
integrable, i.e. that
can be  simultaneously taken to be constant matrices in a suitable coordinate
system,
then   for $n=1$, we have seen that the geometry
 is determined by a 2-form potential, for $n=2$ it is determined by a 1-form
potential, and   that
for the special  case of $n=3$  in which the complex structures satisfy a
quaternion algebra,
by a 0-form potential. In the next section, we will generalise this and show
that
for the general  case of $n=3$ the geometry
 is again determined by a 0-form potential, and for higher $n\ge 3$
it is given by a 0-form potential satisfying certain differential  constraints.

In [\strom], the conditions for
  sigma-models with $N$ supersymmetries
to have superconformal supersymmetry were found for
$N=1$ and $N=2$,  and for $N=4 $ sigma-models with  complex structures
satifying the $SU(2)$ algebra
\quat. These generalise to give  the corresponding conditions for  any $N$, for
an $N$-supersymmetric
 sigma-model to be invariant under the superconformal group $OSp(N/1)$. The
analysis is similar to that in [\strom] and will
be given elsewhere.

An extended superspace form of the sigma-model for general $N$ can be given,
following [\GPS].
Let $X$ be a map
from the  superspace with coordinates $\{t; \theta^0,
\theta^r, r=1,\dots,N-1\}$ into a CKT manifold ${\cal M}$.  Then we
impose the constraints
$$
D_rX^i=I_r{}^i{}_j D_0X^j
\eqn\cseven
$$
where $\{D_0, D_r; r=1,\dots,N-1\}$ are the supersymmetry derivatives
satisfying
$$
\eqalign{
D_0^2&=i{d\over dt}
\cr
D_0D_r+D_r D_0&=0
\cr
D_sD_r+D_r D_s&=2i\delta_{rs}{d\over dt}\ .}
\eqn\abseven
$$
These constraints are consistent provided \clif,\nab\ are satisfied.
An action for this multiplet is
$$
I=-{1\over2}\int\, dt d\theta^0\, \big(i g_{ij}D_0X^i {d\over dt}X^j+
{1\over 3!} c_{ijk} D_0X^i
D_0X^j D_0X^k\big)\ .
\eqn\ceight
$$
and will be independent of $\th ^a$ and hence fully supersymmetric provided
\conn\ are satisfied.

\section{N=2 Superspace Formulation}

The models with $N\ge 2$ supersymmetry can be written in $N=2$ superspace with
action \actt.
It will prove useful to
 rewrite the action using the chiral constraint $\Dd Z=2DZ$ as
$$
I={1\over 4}\int dt\, d^2\theta\, \w_{ij} DX^iDX^j
\eqn\tact
$$
where the
 antisymmetric tensor
$\w_{ij}$
has components
$$
\w_{\a\b}= B_{\a\b}, \qq \w_{\ab\bbb}= \ba B_{\ab\bbb}, \qq
\w_{\a\bbb}=- \w_{\bbb \a}=ig_{\a\bbb}
 \eqn\abc$$
so that the
two-form $\w={1\over 2} \w_{ij}dX^i\wedge dX^j$
is given by
$$\w=\ww +B + \ba B
 \eqn\abc$$
where $B={1\over 2}B_{\alpha\beta}dZ^\alpha
\wedge dZ^j\beta$.
Thus the geometry is specified by the choice of an arbitrary real 2-form $\w$
potential, so that
the model is defined by
a complex manifold with a 2-form, i.e. by the triple $({\cal M}, I, \w)$. The
metric is defined by the
(1,1) part of $\w$ and $B$ by the (2,0) part, and $\w$ is defined up to the
transformations \kagag\
$$\w \to \w + \pa \l + \ba \pa \ba \l
\ek
where $\l$ is an arbitrary (1,0) form.

The action
\tact\ can be expanded
into $N=1$ superspace
to give
$$
I={1\over 4}\int dt\, d \theta\, \left( 3\w_{ijk} DX^iDX^j \hat DX^k
-2i \w_{ij} \hat DX^i \pa _tX^j
\right)
\eqn\actono
$$
where $ \w_{ijk} \equiv \w_{[ij,k]}$,
$$ \hat D \equiv  {1\over 2i} (\Dd-\ba \Dd)
 \eqn\abc$$
and the chiral constraint implies
$$ \hat D X^i = I^i{}_j DX^j
 \eqn\abc$$
Using the notation of  [\pop] that for any tensor
$ T_{ij...kl}$,
$$ T_{ij...k\hat l} = T_{ij...km} I^m{}_l
\eqn\abc$$
the action
\actono\ can be rewritten as
$$
I={1\over 4}\int dt\, d \theta\, \left( 3\w_{ij\hat k} DX^iDX^j  DX^k
-2i \w_{ \hat ij}  DX^i \pa _tX^j
\right)
\eqn\actonor
$$

The fact that, for any tensor $\ss_{ij}$,
$$\int dt\, d \theta\,
  \ss_{ij}   DX^i \pa _tX^j
=
\int dt\, d \theta\, \left(
  \ss_{(ij)}  DX^i \pa _tX^j -{i\over 2}
\ss_{[ij,k]} DX^iDX^jDX^k
\right)
 \eqn\abc$$
can be used with $\ss_{ij}=\w_{ \hat ij}$ to rewrite
the action \actonor\    as
$$
I={1\over 4}\int dt\, d \theta\, \left( [3\w_{ij\hat k}-\w _{\hat i j ,k}]
DX^iDX^j  DX^k
-2i \w_{( \hat ij)}  DX^i \pa _tX^j
\right)
\eqn\actonor
$$
so that, comparing with \act, we have
$$ \eqalign{g_{ij}&= \w_{( \hat ij)}=\w_{k(i}I^k{}_{j)}
\cr
c_{ijk}=&
-3[3\w_{ij\hat k}-\w _{\hat i j ,k}]
=-3 [I{}^l {}_{[i}\w_{jk]l}
+3\pa_{[k}(I{}^l {}_{i}\w_{j]l} ]
\cr}
 \eqn\gcis$$
In form notation, $c$ is
$$
c= -\iota _I d \w + {1\over 2} d \iota _I  \w
 \eqn\abc$$

Finally, note that for $N>2 $ supersymmetry with $N-1$ complex structures
$I_r$, one can choose any
one of them and
  work in the corresponding
$N=2$ superspace, giving a 2-form $\w^r$ for that complex structure, and in
this way one can construct $N-1$ 2-forms
$\w^r_{ij}$.

\chapter{Integrable Complex Structures and Extended Superspace}

\section{ Complex Structures and Clifford Algebras}

In this section we will examine the case in which there are $m$ simultaneously
integrable complex structures $I_a$, each of
which satisfies the conditions \clif,\nab,\conn\ for $N=m+1$ supersymmetry.
In such a case, there
is a coordinate choice in which the $I_a$ are all real constant matrices
satisfying the Clifford
algebra \clif, and the superspace constraints become of the conventional kind.
The superspace
action then  leads to a simplification of the geometry (for example, for $N=4$
supersymmetry, the metric
and torsion are given in terms of a scalar potential $L$).

Real $D\times D$ matrices satisfying \clif\ can be constructed from the basic
real
$2\times 2 $ matrices $\ss_1, \ss_3 , \ee=i\ss_2$.
 For $D=2$, the only  real  matrix satisfying \clif\ is $\ee$, for $D=4$
a set of three matrices $I_r$ ($r=1,2,3$) satisfying \clif\ is given by
$$ \ee \otimes \II, \qq \ss_1 \otimes \ee, \qq \ss_3 \otimes\ee
 \eqn\dfgdfgtj$$
and these in fact satisfy the quaternion algebra \quat.
An alternative set $\ti I_r$ is given by
$$ \II \otimes \ee, \qq \ee \otimes \ss_1, \qq \ee \otimes\ss_3
\ek
and these two sets commute: $[I_r ,\ti I_s]=0$.
For $D=8$, a set of 7 matrices satisfying \clif\ can be constructed from the
$I_r,\ti I_r$:
$$ \ee \otimes \II, \qq \ss_1 \otimes I_r, \qq \ss_3 \otimes \ti I _s
\eqn\ytyrds$$
More generally, if for some $D$ there are two commuting sets of Clifford
structures  $I_r,\ti I_r$,
$r=1,..,m$ for some $m$, a Clifford structure for dimension $2D$ (i.e. with
$2D\times 2D$ matrices)
is given by the $2m+1$ matrices \ytyrds.

  \section{ N=3  Supersymmetry}

In this case there are simultaneously integrable  complex structures $I,J$ and
the superspace constraints are
$$ D_1 X^i= I^i{}_j DX^j, \qq
 D_2 X^i= J^i{}_j DX^j
\eqn\dsada$$
with $D=D_0$.
The product $\ti K\equiv IJ$ is also a complex structure (as $IJ=-JI$), and
$I,J,\ti K$   satisfy  the algebra of unit
imaginary quaternions and are simultaneously integrable. In general $\ti K$
will not satisfy the conditions \conn\
so that the action will  be invariant under the supersymmetry transformations
corresponding to $I$ and $J$ but not $\ti K$.

The general form of the $N=3$ action (with simultaneously integrable  complex
structures) is given in terms of in terms of an
arbitrary 1-form potential $k$ by
\acth, which can be rewritten  as
$$I= {1\over 4} \int dt\, d \theta ^0 d \theta^ 1 d \theta^2 \, k_i DX^i
\eqn\iksi$$
using the constraints \dsada.
Performing the $\th ^2$ integral gives the $N=2$ superspace action
$$I={1\over 2} \int dt\,d \theta^0 d \theta ^1    \,  k_{[j,k]}J^k{}_iDX^iDX^j
 \eqn\abc$$
so comparing with
\tact\ gives
$$\w _{ij}=  J^k{}_{[i}k_{j]k}
\eqn\abc$$
where $k_{ij}=2k_{[i,j]}$,
or
$$
\w = 2 \i_J dk
\eqn\abc$$
The metric and torsion are then given by \gcis, so that
$$\gij
=    k_{k(i}\ti K ^k_{j)} + I^k{}_{(i} J^l{}_{j)} k_{kl}
\ek
The condition for the complex structure $\ti K$   to give  a fourth
supersymmetry is that
$$d \iota _{\ti K} k=0
\ek
so that locally
there is a scalar $L$ such that
$$ k_i = \ti K ^j{}_i \pa _j L
\ek
as will be seen in the next section.

 \section{ N=4  Supersymmetry}

In this case there are simultaneously integrable  complex structures $I,J ,K$
and the superspace constraints are
$$ D_1 X^i= I^i{}_j DX^j, \qq
 D_2 X^i= J^i{}_j DX^j, \qq
D_3 X^i= K^i{}_j DX^j
\eqn\abc$$
The case in which  $IJ =K$ and $I,J ,K$  satisfy  the algebra of unit
imaginary quaternions has been analysed in section 4.3. Here we will not assume
this, so that in general
there are three additional   complex structures defined by the products
$$\tilde K=IJ, \qquad \tilde I= JK , \qquad \tilde J= KI
\ek
and there is a real structure ($R^2=\II$)
$$R=IJK
\ek
If the complex structures satisfy \quat, the dimension is $D=4n$ for some $n$
and the complex structures can be taken to be
$I_r  \times \II_n$
where $I_r$ are the $4\times 4 $ real matrices \dfgdfgtj\ and $\II_n$ is the
$n \times n $ identity matrix.
If the complex structures do not satisfy \quat, the dimension is $D=8n$ for
some $n$
and the complex structures can be taken to be
$$ \ee \otimes \II\otimes \II, \qq \ss_1 \otimes \ee \otimes \II, \qq \ss_3
\otimes \II
\otimes\ee
 \eqn\dfgdsdffgtj$$

In the coordinate system in which the complex structures are all constant, the
general form of the $N=4$ superspace action
is
$$I= {1\over 4} \int dt\, d \theta ^0 d \theta^ 1 d \theta^2 d \theta^3 \,L(X)
\eqn\actff$$
for some potential $L$.
Performing the $\th ^3$ integral gives the $N=3$ superspace action
$$I={1\over 4} \int dt\, d \theta ^0 d \theta^ 1 d \theta^2
 \,  L,_j K^j{}_iDX^i
 \eqn\abc$$
which is of the same form as \iksi, with the 1-form potential given by
$$k_i=L,_j K^j{}_i
\eqn\abc$$
or
$$k=\i _{K} dL
\eqn\abc$$
so that the geometry 2-form is given by
$$
\w = 2 \i_{J} d(\i _{K} dL)
\eqn\abc$$
Then the metric and torsion are given by \gcis.
Defining
$$
P^{ij}{}_{kl}= \tilde I ^i{}_k I^j{}_l +
 \tilde J ^i{}_k J^j{}_l
+
 \tilde K ^i{}_k K^j{}_l
\ek
this gives
$$g_{jk}=  {1 \over 2} L,_{ml}
P^{ml}{}_{(kj)} -{1 \over 2}
R^l{}_{(j} \pa _{k)} \pa _l L
\ek
and
$$I_{[j}{}^iB_{k]i}= {1 \over 2} L,_{ml}
P^{ml}{}_{[kj]} -{1 \over 2}
R^l{}_{[j} \pa _{k]} \pa _l L \ek
In the special case of a quaternionic structure with $I=\tilde I, J=\tilde J,
K=\tilde K$ and $R=-\II$, then the space is weak HKT,
$B=0$ and the expression for the metric is given by
$$g_{kl}= {1 \over 2} \left(  L, _{kl}
+
 [ I ^i{}_k I^j{}_l +
   J ^i{}_k J^j{}_l
+
  K ^i{}_k K^j{}_l ]
L,_{ij} \right)
\ek
as in [\strom], and agreeing with
the results of section 4.3.

\section{N>4 Supersymmetry}

For any model with $N\ge 4 $ supersymmetry and simultaneously integrable
complex structures, the
action can be written in  the  $N=4$ superspace corresponding to {\it any}
 three complex structures  in terms of a scalar
potential (with different potentials for different sets of three).
If any three of the  complex structures  satisfy the quaternion algebra \quat,
then the
geometry must be weak HKT and there can be no more than 4 supersymmetries
unless the holonomy of the connection
$\ggg ^{(+)}$
is trivial.
For $N>4$, the action can be written in full extended superspace in  a same way
similar to that used in [\gates], giving
an explicit expression for the  potential $L$ as a multiple contour integral.
Alternatively, the
conditions for the $N=4$ action to have further supersymmetries
leads to differential constraints on $L$ (similar to those in [\gates], the
general solution of which
is given by the multiple contour integral expression. Details will be given
elsewhere.

\ack
{I would like to thank George Papadopoulos for helpful discussions.}

\appendix


{For any  $N=2$ superspace action of the form
$$I= \int dt\, d \theta ^1 d \theta^2 \, L
 \eqn\abc$$
the $\th^2 $ integral gives
$$I= \int dt\, d\theta\, {1\over 2i} (\Dd-\ba \Dd) L
 \eqn\abc$$
(Recall that $\Dd=D_0+iD_1$ and $D=D_1$.)
For
$$L=iG_{\a \bbb} \Dd Z^\a
\ba \Dd \ba Z^{\bbb}
 \eqn\abc$$
this gives
$$\eqalign{
{1\over 2i} (\Dd-\ba \Dd) L
&=-2i
G_{\a \bbb}( D Z^\a
 \pa _t \ba Z^{\bbb}
+\pa _t Z^\a
 D \ba Z^{\bbb})
\cr &
-4G_{\a \bbb,\gg } D Z^\a D Z^\gg D \ba Z^\bbb
 -4G_{\ab \b,\ggb } D \ba Z^\ab D \ba  Z^\ggb D   Z^\b
\cr}
 \eqn\abc$$
where the chiral  constraint $\Dd Z=2DZ$ has been used.
For
$$L={1\over 2} B_{\a \b}\Dd Z^\a \Dd Z^\b
 \eqn\abc$$
it is useful to write
$$ (\Dd-\ba \Dd) L= 2\Dd L -2D L
 \eqn\abc$$
to obtain
$${1\over 2i} (\Dd-\ba \Dd) L
=-4i B_{[\a \b , \gg ]}DZ^\a DZ^\b DZ^\gg
+i DL
 \eqn\abc$$
and the term $iDL$ is a surface term in the superspace action, which can be
discarded.}


\refout
\bye